\documentstyle[sprocl]{article}
\input{epsf}
\def\Journal#1#2#3#4{{#1} {\bf #2}, #3 (#4)}

\def\NPB{{\em Nucl. Phys.} B}
\def\PLB{{\em Phys. Lett.}  B}
\def\PRL{\em Phys. Rev. Lett.}
\def\PRD{{\em Phys. Rev.} D}
\def\ZPC{{\em Z. Phys.} C}

\def\mco{\multicolumn}

\def\ra{\rightarrow}

\def\be{\begin{equation}}
\def\ee{\end{equation}}
\def\bea{\begin{eqnarray}}
\def\eea{\end{eqnarray}}

\def\ghad{\Gamma_{\rm had}}
\def\gb{\Gamma_{\rm b}}
\def\etal{{\it et.al.}}
\def\zo{Z^o}
\def\dg{\delta g}
\def\afbt{$A_{\rm FB}^o(\tau )$}
\def\alr{$A_{\rm LR}^o$}
\def\vtb{{\cal V}_{tb}}
\def\vtpb{{\cal V}_{t'b}}
\def\i3p{{I'_3}}
\def\roughly#1{\mathrel{\raise.3ex\hbox{$#1$\kern-.75em\lower1ex
\hbox{$\sim$}}}}

\begin{document}
\hfill McGill/96-21 \\
\title{$R_b$ AND HEAVY QUARK MIXING}
\author{P. BAMERT}
\address{Physics Department, McGill University, 3600 University St.,\\
Montr\'eal, Qu\'ebec, Canada, H3A 2T8}
\maketitle\abstracts{In this talk I summarize a part of the work 
done in a recent collaboration with C. Burgess, J. Cline, D. London and 
E. Nardi \cite{b96}. We analyze the observed discrepancy 
between $R_b(\equiv\gb /\ghad )$
as measured at LEP and its standard model value for signals of
new physics. The focus is thereby put on new physics that manifests itself
through heavy quark mixing. Heavy quark mixing affects the measured value
of $R_b$ in two ways: at tree level (bottom mixing) and at one-loop level 
(top mixing). One finds that whereas the latter cannot account for the 
deviation, bottom mixing can in principle do the job.}
During the past years measurements of $e^+e^-$ scattering on the $\zo$ 
resonance at LEP \cite{lep} and SLC \cite{slc} have confirmed the standard 
model (SM) of electroweak interactions to astounding levels of precision, 
culminating in a prediction of the top mass in agreement with actual 
observations at CDF and D0 \cite{cdf}. Despite this success the SM is 
now excluded at the $97.5\% $ C.L. {\it if} one takes the data at face value.
The culprit is the, by now well known, observed surplus of bottom quarks
produced in $\zo$ decays. The relevant observable, $R_b$, deviates from its
SM value by $3.4\sigma$. In fact one has \cite{lep}
\be 
R_b \equiv {\Gamma (\zo\ra b\bar{b})\over\Gamma (\zo\ra{\rm hadrons})}=
0.2211\pm 0.0016\quad{\rm (EXP)}\qquad 0.2156\quad{\rm (SM).}
\label{eqrb} 
\ee
The aim of this talk is to summarize 
some of the implications $R_b$ has for new physics. We thereby focus
on bottom- and top-mixing as potential explanations of $R_b$.
Other scenarios such as SUSY and generic scalar-fermion loop corrections
to the $\zo b\bar{b}$ vertex are discussed elsewhere in these 
proceedings \cite{c96}, by J. Cline. The two talks add up to summarize the 
essential contents of a paper done in collaboration with
C.P. Burgess, J. Cline, D. London and E. Nardi \cite{b96}.

The basic philosophy pursued here is to be as unbiased as possible,
as far as the actual field content of the added new physics is concerned, 
in order to identify the mechanisms that can bring about the needed 
variation in $R_b$. 

The next section will briefly review the experimental situation while
sections 3 and 4 will discuss bottom- and top-quark mixing respectively.

\section{The Experimental Situation}
In order to see what actually needs explaining we parametrize the indirect
effects of new physics in terms of an effective lagrangian \cite{b94}.
In the case of $\zo$-pole measurements it turns out that it actually suffices
to consider only terms up to dimension four. These describe 
modifications to the neutral current couplings of the 
fermions ($\dg_{L,R}^f$) as well as to the gauge
boson vacuum polarizations (through the Peskin-Takeuchi parameters 
$S$ and $T$ \cite{p90}). We normalize these parameters such that
\be
{\cal L}_{\rm eff}^{\rm nc} = {e\over s_w c_w} \, Z_\mu \overline{f} 
\gamma^\mu \left[ \left( g_L^f + \dg_L^f \right) \gamma_L +
\left(g_R^f + \dg_R^f \right) \gamma_R \right] f .
\label{eqefflag}
\ee
so that the SM couplings $g_{L,R}^f$ can be written in terms of the third
component of the weak isospin $I^{3,f}_{L,R}$ and the electric charge $Q^f$
as $g_{L,R}^f = I^{3,f}_{L,R} - Q^fs_w^2$. Here $s_w$ and $c_w$ denote the 
sine and cosine of the weak mixing angle respectively. Notice also that 
$I^3_R=0$ for SM fermions.  

Fitting these new physics parameters to the experimental results one 
learns that it is necessary and sufficient to modify only the $\zo b\bar{b}$
couplings $\dg_{L,R}^b$ \cite{b95}. In other words it suffices to 
explain $R_b$. Deviations from the SM observed in other observables, such
as $R_c$, \afbt\ and \alr, all can be viewed as statistical fluctuations.
This is in part because they only deviate at the $\sim 2\sigma$ level - 
something that could well arise statistically given the numbers of 
observables~\footnote{Notice that $R_c$, according to the latest data 
update \protect{\cite{lep}}, now merely deviates by $-1.8 \sigma$.} - and in
part because they are hard to account for by new physics that manifests itself
through the effective lagrangian of Eq. \ref{eqefflag} \cite{b95}.

\begin{figure}
\begin{center}
\leavevmode
\hbox{%
\epsfxsize=2.8in
\epsfysize=2.45in
\epsfbox{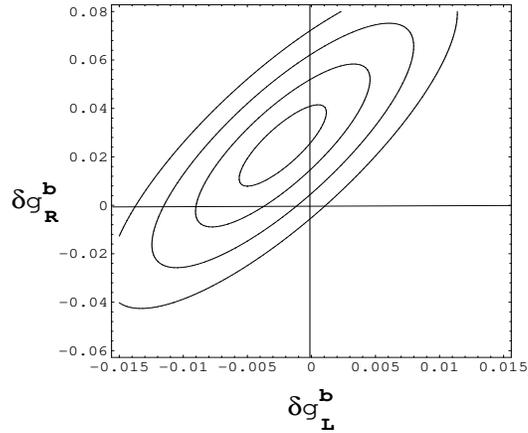}
}
\end{center}
\caption{A global fit of both $\dg_L^b$ and $\dg_R^b$ to the 
Moriond 1996 $\zo$-pole data \protect\cite{lep}. 
The four solid lines respectively 
denote the 1-, 2-, 3- and the 4-$\sigma$ error ellipsoids. Clearly there is
no significant preference for a correction to the left- vs. a 
correction to the 
right-handed $\zo b\bar{b}$ coupling. 
This fit yields a low value for the strong
coupling constant, $\alpha_s=0.102\pm 0.007$, in agreement with low-energy
determinations.} 
\label{figbfit}
\end{figure}

The results of a fit that treats $\dg_{L,R}^b$ as free parameters are shown 
in Figure~\ref{figbfit} as well as in Table~\ref{tabbfit}. The $\chi^2/
{\rm d.o.f.}$ and confidence levels of the two individual fits shown in 
the table are respectively $14.7/12$ ($26\%$ C.L.), for $\dg_L^b$, and
$13.0/12$ ($37\%$ C.L.), for $\dg_R^b$. This is to be compared with the
standard model for which $\chi^2/{\rm d.o.f.}$ is $24.7/13$ ($2.5\%$ C.L.).
\begin{table}[t]
\caption{Individual fits of separately $\dg_L^b$ and $\dg_R^b$
to Moriond 1996 data \protect\cite{lep}, 
compared to the SM tree level values for
the $\zo b\bar{b}$ couplings and to the SM dominant $m_t$-dependent one-loop
vertex correction. The latter has been evaluated at $s_w^2=0.23$ and $m_t=180$
GeV.\label{tabbfit}}
\vspace{0.4cm}
\begin{center}
\begin{tabular}{|c|c|c|c|}
\hline
Coupling & $g$(SM tree)&$\dg$(SM top loop)&$\dg$(Individual Fit)\\
$\dg_L^b$&$-0.4230$&$0.0065$&$-0.0063\pm0.0020$\\
$\dg_R^b$&$0.0770$&$0$&      $ 0.034 \pm0.010$\\      
\hline \end{tabular} \end{center}
\end{table}
Notice that these fits not only 'explain' $R_b$ but at the same time also
point to a low value for the strong coupling constant $\alpha_s$ which is
in better agreement with low-energy determinations than the value obtained
from a SM fit \cite{h94,b95}.

Figure \ref{figbfit} and Table \ref{tabbfit} contain more information
that helps us pinpoint what kinds of new physics we actually need in order to 
explain $R_b$ (and with it the data). These are:
\begin{itemize}
\item
There is no statistically significant preference for a new physics correction
to the left-handed vs. a correction to the 
right-handed $\zo b\bar{b}$ coupling.\footnote{This can be seen either 
from the strong correlation between $\dg_L^b$ and $\dg_R^b$ as 
reflected in the tilted ellipsoid in Figure \ref{figbfit} or 
from the fact that the two individual fits of Table \ref{tabbfit} 
both have high confidence levels compared to the SM.}
\item
$\dg_L^b$ looks to be the size of a {\em largish loop}, albeit with 
a sign opposite to the SM one-loop top quark correction.
\item 
$\dg_R^b$ on the other hand looks more like a {\em tree level effect} since
it appears to be too big to be accounted for by a loop correction.
\end{itemize}

Finally a word on the gauge boson vacuum polarizations ($S$ and $T$). 
Since they
represent universal corrections to all $\zo$-fermion couplings they cancel to
a large degree in $R_b$ and are therefore not directly relevant for $R_b$.
Still, any model of new physics emploied to explain $R_b$ must respect the
bounds put on $S$ and $T$ by the other LEP/SLC observables \cite{b95}.

\section{Bottom Mixing}
Without much loss of generality it suffices to consider the case
where the SM $b$-quark mixes with only one new $b'$-quark \cite{b96}.
To fix the notation let us denote the flavour eigenstates with $B$ and $B'$
and the mass eigenstates with $b$ and $b'$. The $B$-$B'$ sector then 
sports a $2\times 2$ mass matrix which in general isn't symmetric: 
\be
\left(\bar{B}\quad\bar{B'}\right)_L\left(\begin{array}{cc}M_{11}&M_{12}\\
M_{21}&M_{22}\end{array}\right)\left(\begin{array}{c}B\\B'\end{array}
\right)_R
\label{eqbmass}
\ee
To diagonalize this matrix one has to rotate the left- and right-handed 
fields separately. 
Let us denote the two corresponding mixing 
angles by their sine (and cosine) $s_{L,R}$ ($c_{L,R}$). Assuming the $b'$
to be too heavy to be directly produced, the mixing then modifies the
tree level $\zo b\bar{b}$ couplings to become
\be
g_{L,R}^b=g^B_{L,R}c_{L,R}^2+g^{B'}_{L,R}s_{L,R}^2
\label{eqmixcoup}
\ee
In terms of the third component of the weak isospin $\i3p_{L,R}$ of the
$B'$-quark, and neglecting $m_b$, the $b$-width is proportional to
\be
\Gamma_b\propto (g_L^b)^2+(g_R^b)^2=\left(
-{c_L^2\over2}+{s_w^2\over 3}+s_L^2\i3p_L\right)^2+
\left({s_w^2\over 3}+s_R^2\i3p_R\right)^2
\label{eqmixwidth}
\ee
To increase $R_b$ in magnitude one therefore either needs to decrease
$g_L^b$ and/or to increase $g_R^b$. This requirement then leads to the 
following conditions on the third components of the $B'$ weak isospins:
\be
\begin{array}{ccc}{\rm Small\  mixing}&\qquad&{\rm Large\ mixing}\\
\i3p_L<-{1\over 2}\quad{\rm or}\quad \i3p_R>0&\qquad&
\i3p_L>0 \quad {\rm or}\quad \i3p_R<0
\end{array}
\label{eqi3cond}
\ee   
Large mixing here means that $s_L^2(\i3p_L+{1\over 2})>1-2s_w^2/3\sim0.85$
or $s_R^2|\i3p_R|>2s_w^2/3\sim 0.15$ in order to actually increase $R_b$.

Notice that the presence of left-handed mixing modifies the CKM matrix element
$\vtb$ to become $c_L\vtb$. Surprisingly a large left-handed mixing is at 
present not experimentally excluded \cite{l96} since CDF and 
D0 do not directly measure $\Gamma(t\rightarrow bW)$ (and hence $\vtb$), rather
they constrain the branching ratio $\Gamma(t\rightarrow bW)/\Gamma(t
\rightarrow qW)$.

\begin{table}[t]
\caption{All possible models of $b$-mixing that can explain $R_b$ ({\it i.e.}
fulfill Eq. \protect{\ref{eqi3cond}}) subject to the assumptions that there
are no Higgs representations beyond doublets and singlets and that the $B$ 
mixes with a single $B'$ only. In the two left-most columns the weak isospin
assignements of the $B'$ are given. The last column shows which mixings
(left- 'L' and/or right-handed 'R') affect $R_b$ (Subleading mixings, 
quadratically suppressed, are indicated by brackets) as well as the required
size of the mixing angles.
\label{tabbsol}}
\vspace{0.4cm}
\begin{center}
\begin{tabular}{|c|c|c|c|c|}
\hline
$(I',\i3p )_L$&$(I',\i3p )_R$&Model&\mco{2}{|c|}{Required Mixing}\\
\hline
$(1,-1)$&$(1,-1)$&Vector Triplet \cite{m96}&L,(R)&  
$s_L^2=0.0111\pm 0.0032$\\
$(1,-1)$&$({1\over 2},-{1\over 2})$&&L,(R)&dito\\
$({3\over 2},-{3\over 2})$&$(1,-1)$&&L,(R)& $s_L^2=0.0056\pm 0.0016$\\
\hline
$({1\over 2},{1\over 2})$&$({1\over 2},{1\over 2})$&Vector Doublet \cite{ch6}
&(L),R& $s_R^2=0.052 {+ 0.013\atop -0.014}$\\
$(0,0)$&$({1\over 2},{1\over 2})$&&(L),R&dito\\
$({1\over 2},{1\over 2})$&$(1,1)$&&(L),R& $s_R^2=0.026 {+ 0.006\atop -0.007}$\\
\hline
$({1\over 2},{1\over 2})$&$(0,0)$&&L& $s_L^2=0.8515\pm 0.0016$\\
$({1\over 2},{1\over 2})$&$(1,0)$&&L&dito\\
$({3\over 2},{1\over 2})$&$(1,0)$&&L&dito\\
\hline
$(0,0)$&$({1\over 2},-{1\over 2})$&Mirror Family
&L,R& $s_R^2{\roughly >} 0.361$\\
$({1\over 2},-{1\over 2})$&$({1\over 2},-{1\over 2})$&Vector Doublet 
\cite{y95}&R&$s_R^2=0.361 {+ 0.013\atop -0.014}$ \\
$({1\over 2},-{1\over 2})$&$(1,-1)$&&R&$s_R^2=0.180\pm 0.007$\\
\hline \end{tabular} \end{center}
\end{table}

In order to find all possible solutions to Eq. \ref{eqi3cond} one 
simply starts out by enumerating all weak representations $B'$ can 
have \cite{b96} while requireing
\begin{itemize}
\item at least one non-zero off-diagonal element in Eq. \ref{eqbmass},
{\it i.e.} mixing,
\item a heavy $b'$, {\it i.e.} $M_{22}\ne 0$,
\item no Higgs representations beyond doublets and singlets.
\end{itemize}
Notice that the third requirement does not severely impair the 
generality of the analysis.\footnote{Higher 
dimensional representation would spell trouble for
the $\rho$-parameter if they were to contribute significantly to the $B$-$B'$
mass matrix.}
Not all of the $B'$ representations so obtained fulfill
Eq. \ref{eqi3cond} ({\it i.e.} are able to explain $R_b$). 
In Table \ref{tabbsol} we list only those which do. Some of the options
listed there have been discussed in detail in the literature 
\cite{m96,ch6,y95}. 
Notice that not all of the $B'$ representations listed in Table \ref{tabbsol} 
are anomaly free per se. In these cases it is however 
straightforward to add other fermions, which don't affect $R_b$, 
in order to cancel the anomaly.

Two things we notice from looking over Table \ref{tabbsol}. First some 
popular and simple extensions of the SM are not present. This is because
these either decrease $R_b$ (as does the Vector Singlet model
with $I'_L=I'_R=0$) or don't affect this observable at all (as is the case 
for a fourth family). Second, some of the third weak isospin components of
the models listed seem to have the wrong value for increasing $R_b$. 
In most of these cases the corresponding mixing angle is however suppressed,
because gauge invariance requires one of the off-diagonal mass matrix 
elements of Eq. \ref{eqbmass} to be zero.
  
\section{Top Mixing}
Top quark mixing, being a one-loop effect, can never produce
a right-handed $\zo b\bar{b}$ vertex correction large enough to explain
$R_b$. As we see from Table \ref{tabbfit} however, the required value
for $\dg_L^b$ is about equal in magnitude and opposite in sign to the 
SM top quark correction. Alas, there seems to be some hope that mixing
effects could somehow reverse the sign while the large top-quark Yukawa
couplings provide the needed magnitude of the correction. 
As it turns out and as we will explain a bit more in detail 
below this is, however, not the case.

In a way very analogous to the one described in the previous section one
can list all possible models of $t$-mixing \cite{b96}. This is done 
under essentially identical assumptions, {\it i.e.} no higher dimensional
Higgs representations and the mass matrix has to exhibit both mixing and
a mass for the $T'$. A new aspect arises from the presence of a $B'$ in 
many of these models since tree level $b$-mixing could potentially
dominate any loop induced corrections due to $t$-mixing. According to the
nature of the involved $B'$, models for $t$-mixing fall into four 
categories:
\begin{itemize}
\item those in which the $B'$ is SM like, {\it i.e.} has the same quantum 
numbers as the SM $B$, and hence does not affect $R_b$,
\item those in which the $B'$ is exotic, {\it i.e. not} SM like, but in which
gauge invariance imposes a constraint on the $B$-$B'$ mass matrix that forbids
$b$-mixing,
\item and those in which the $B'$ is exotic {\it and} mixes, in which case 
we impose the additional constraint that $b$-mixing vanishes in order to 
isolate the loop-effect.
\item Finally there are those models that do not contain a $B'$.
\end{itemize}
For a detailed discussion and complete list of models so obtained refer to
the paper on which this talk is based \cite{b96}.

\begin{figure}
\begin{center}
\leavevmode
\hbox{%
\epsfxsize=3.9in
\epsfbox{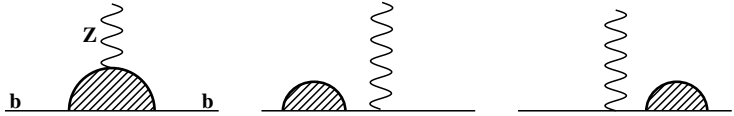}
}
\end{center}
\caption{General Feynman diagrams relevant for $t$-mixing corrections to $R_b$.
The direct vertex correction and the b-quark self-energies sum up to yield a
UV finite result.
} 
\label{figvc}
\end{figure}
Before even starting to
compute the correction to the left-handed $\zo b\bar{b}$ vertex 
we can infer some of its
properties from gauge invariance. If the external gauge boson 
was a photon (and not a $\zo$) then the electron self energies and the 
vertex correction shown in Figure \ref{figvc} would exactly cancel at
zero momentum transfer ($q^2=0$) as a consequence of the electromagnetic 
Ward Identity. With broken gauge symmetry and the external gauge boson being
a $\zo$ those terms of the correction that remain unchanged still cancel.
Specifically these are the divergencies ({\it i.e.} the total correction
to $\dg_L^b$ is UV finite) and the terms proportional to the coupling of
the photon ({\it i.e.} the electric charge $Q$ and hence $s_w^2$). Away from
zero momentum transfer the latter no longer holds, the corresponding 
correction however turns out to be small. In the SM case one obtains
(in the limit of a large top mass, $r \gg  1$): \cite{b88}
\be
{\dg_L^b}^{\rm SM}\approx {\alpha\over 16\pi s_w^2}\left[r +\left( 3-{s\over 6}
(1-2s_w^2)\right)\ln r\right]
\label{eqsmcorr}
\ee
where $r=m_t^2/M_W^2$, $s=q^2/M_W^2=M_Z^2/M_W^2$ and $\alpha$ is the 
fine structure constant. As stated above the term 
proportional to $s$ is numerically small.

Turning now to the case of $t$-mixing what has been said above still 
holds. Specifically, we can neglect terms proportional to $s$, as they 
turn out to be irrelevant for the precision needed in the
present analysis. Also, although the final result is more complicated, the
total correction is still UV finite and independent of electric quark charges.

To diagonalize the $T$-$T'$ mass matrix we rotate, as before, the left- and
right-handed chiralities separately. These rotations modify the neutral
current couplings of the top quarks:
\be
g_{L,R}^{ij}=\sum_{a=T,T'}g_{L,R}^a{\cal U}_{L,R}^{ai}{\cal U}_{L,R}^{aj}
\label{eqncc}
\ee
where ${\cal U}$ denotes a two-by-two rotation matrix and $i,j=t,t'$. 
Clearly $t$-mixing introduces
flavor changing neutral currents that are of relevance in vertex correction
diagrams. The mixing also affects the charged current couplings. In the
presence of $b$-mixing one then has $\vtb=c_Lc_L^B+s_Ls_L^B$ and 
$\vtpb=s_Lc_L^b+c_Ls_L^B$. Here the superscript $^B$ indicates the b-quark
mixing angles. Defining $r'\equiv m_{t'}^2/M_W^2$ the total correction,
$\dg_L^b$, can be viewed as a dimensionless, lorentz-invariant form factor
which depends on $r$,$r'$, the weak isospin assignements of the $T'$ field and
the mixing angles. Defining the 'net' correction due to $t$-mixing as the
difference between the total and the SM top-quark correction, 
$\dg_L^b={\dg_L^b}^{\rm total}-{\dg_L^b}^{\rm SM}$, one then obtains
\cite{b96}
\bea
\dg_L^b &=& {\alpha\over 16\pi s_w^2}\left\{
\vtpb^2\left[ r'-r+3\ln \left({r'\over r}\right)\right]\right.\nonumber\\
&&\qquad +(1-2\i3p_L)\vtb\vtpb s_Lc_L\left[-r-r'+{2rr'\over r'-r}\ln\left(
{r'\over r}\right)\right]\label{eqtmixcorr}\\
&&\qquad +2\i3p_R\vtb^2s_R^2\left[-r+{1\over 2}\left(1+{r\over r'}
\right){rr'\over r'-r}\ln\left({r'\over r}\right)\right.\nonumber\\
&&\qquad\qquad\qquad\qquad \left.\left. -{3r\over r'-r}
\ln\left({r'\over r}\right) + {3\over 2}\left( 1+{r\over r'}\right)\right]
\right\}\nonumber
\eea
which is valid in the limit of heavy quark masses 
($r,r'\gg 1$).
Upon analyzing this result more in detail one finds that even for $m_{t'}>
m_t$ it is possible to choose $\i3p_{L,R}$ and mixing angles such that the
correction is negative~\cite{b96}. So $t$-mixing can indeed increase $R_b$. 
To see how large an increase we can, in principle, get, we choose the most 
optimal values for the parameters that control Eq. \ref{eqtmixcorr}: 
$\vtpb\approx 1$ and $m_{t'}\approx  135$ GeV. In this case the first term
of Eq. \ref{eqtmixcorr} dominates and one obtains, for $m_t=180$ GeV,
$\dg_L^b\approx -0.0021$
which, according to Table~\ref{tabbfit}, is about a third of what is needed.

\section{Conclusions}
In summary I have argued that in order to explain the current experimental
situation it is necessary and sufficient to modify the theoretical prediction
of $R_b$. Other deviations from the SM can well be viewed as statistical 
fluctuations. The needed correction to the $\zo b\bar{b}$ coupling
is small, the size of a large one-loop effect, if it were left-handed, or 
large, corresponding to a tree-level effect, if it were right-handed.
In the framework of heavy quark mixing we found that $b$-mixing can indeed
explain $R_b$ - provided one is 'exotic' enough. Top quark mixing on the
other hand can modify the radiative corrections to the $\zo b\bar{b}$ vertex
only to the extend of reducing the discrepancy to the data by one third at
best.

\section*{Acknowledgements}
The author would like to thank Cliff Burgess, Jim Cline, David London
and Enrico Nardi for the fruitful collaboration \cite{b96} on which this talk
is based. This research was financially supported by NSERC of Canada and 
FCAR du Qu\'ebec.

\end{document}